\documentclass{moriond}

\def\Journal#1#2#3#4{{#1} {\bf #2}, #3 (#4)}

\def\PRL{\em Phys. Rev. Lett.}
\def\PRD{{\em Phys. Rev.} D}

\def\be{\begin{equation}}
\def\ee{\end{equation}}
\def\bea{\begin{eqnarray}}
\def\eea{\end{eqnarray}}

\usepackage{bm}
\newcommand*{\doi}[2]{\href{http://dx.doi.org/#1}{#2}}

\newcommand{\Photo}{}

\begin{document}
\vspace*{4cm}
\title{VIOLATION OF THE EQUIVALENCE PRINCIPLE FROM LIGHT SCALAR FIELDS: FROM DARK MATTER CANDIDATES TO SCALARIZED BLACK HOLES}

\author{A. HEES$^{1}$, O. MINAZZOLI$^{2,3}$, E. SAVALLE$^{1}$, Y. V. STADNIK$^4$, P. WOLF$^{1}$, B. ROBERTS$^1$}

\address{$^{1}$SYRTE, Observatoire de Paris, Universit\'e PSL, CNRS, Sorbonne Universit\'e, LNE, 61 avenue de l'Observatoire 75014 Paris, France \\ 
$^2$ Centre Scientifique de Monaco, 8 Quai Antoine 1er, 98000 Monaco, Monaco \\ $^3$ Artemis, Universit\'e C\^ote d'Azur, CNRS, Observatoire C\^ote d'Azur, BP4229, 06304, Nice, France \\
$^4$ Helmholtz Institute Mainz, Johannes Gutenberg University, 55099 Mainz, Germany}

\maketitle\abstracts{
Tensor-scalar theory is a wide class of alternative theory of gravitation that can be motivated by higher dimensional theories,  by  models of dark matter or dark ernergy. In the general case, the scalar field will couple non-universally to matter producing a violation of the equivalence principle. In this communication, we review a microscopic model of scalar/matter coupling and its observable consequences in terms of universality of free fall, of frequencies comparison and of redshifts tests. We then focus on two models: (i) a model of ultralight scalar dark matter and (ii) a model of scalarized black hole in our Galactic Center. For both these models, we present constraints using  recent measurements: atomic clocks comparisons, universality of free fall measurements, measurement of the relativistic redshift with the short period star S0-2 orbiting the supermassive black hole in our Galactic Center.}

\section{Introduction}
The theory of General Relativity (GR) is the current paradigm to describe the gravitational interaction. Since its creation in 1915, GR has been confirmed by experimental observations (e.g. Will~\cite{will}). Although very successful so far, it is nowadays commonly admitted that GR is not the ultimate theory of gravitation. Attempts to develop a quantum theory of gravitation or to unify gravitation with other fundamental interactions lead to deviations from GR. Moreover, observations requiring the introduction of dark matter (DM) and dark energy are sometimes interpreted as a hint that gravitation presents some deviations from GR at large scales.

Tensor-scalar theories consist in a large class of widely studied extensions to GR. In these alternative theories of gravitation, one introduces a scalar field $\varphi$ in addition to the standard space-time metric $g_{\mu\nu}$ to describe the gravitational interaction. These theories have been motivated because scalar fields arise naturally in higher dimensions theories, in massive gravity, in string theories and in some models of DM and dark energy. 

In the most general case, the scalar field can couple non-minimally to standard matter and produces a violation of the Einstein Equivalence Principle (EEP) characterized by a violation of the universality of free fall (UFF) and of the local position and Lorentz invariance~\cite{will}. In Section~\ref{sec:theory}, we will review how a non-minimal scalar/matter coupling will impact the motion of test masses, the comparison of frequencies and measurements of the gravitational redshift.

Then, we will focus on two specific models of tensor-scalar theory. The first model, presented in Section~\ref{sec:DM}, is an ultralight scalar DM candidate, an alternative to standard fermionic DM that has not been detected with particles accelerators so far. This bosonic DM candidate can be searched for using atomic clocks comparisons and UFF measurements. We review the constraints on this model available with the state-of-the art current measurements.
 
The second tensor-scalar theory considered in this communication is the quadratic Einstein-Gauss-Bonnet theory presented in Section~\ref{sec:EGB}. In this class of theory, a coupling between the scalar field and the Gauss-Bonnet curvature invariant is introduced motivated by string theory. Such a coupling can give rise to scalarized black holes (BH), i.e. to BH with a large scalar hair. We use this model as a testbed alternative for Sagitarius A$^*$ (Sgr A$^*$), the supermassive black hole (SMBH) in our Galactic Center (GC). We show that recent measurements of the gravitational redshift of the short-period star S0-2 orbiting Sgr A$^*$ can be used to search for a violation of the EEP for a model where Sgr A$^*$ is a scalarized black hole.

\section{Violation of the Einstein Equivalence Principle due to a scalar field: microscopic modeling and observable effects}\label{sec:theory}
The violation of the EEP is totally encoded in the matter part of the action. In this section, we give an example of a Lagrangian describing the interaction between the scalar field and standard matter that explicitly breaks the EEP. In addition, we  give the expression of three observables that are related to a violation of the EEP: (i) the Universality of Free Fall measurements, (ii) the searches for variations of the constants of Nature using atomic sensors or spectral lines and (iii) gravitational redshift tests. The expressions of the observables presented in this section are general and totally independent of the full action (in particular, they are independent of the kinetic and potential part for the scalar field) and of the interpretation of the scalar field (as DM, dak energy or anything else).

\subsection{Microscopic modeling}\label{sec:micro}
One way to break the EEP is to introduce a non-minimal coupling between the scalar field and the fields from the standard model (SM) of particle physics. By non-minimal coupling, we mean a coupling that can not be reabsorbed by a global conformal or disformal transformation. The number of ways to non-minimally couple a scalar field to standard matter is only limited by our imagination and several models have been studied in the literature like e.g.: introduce different conformal couplings between the scalar field and different parts of the SM Lagrangian (see e.g. Khoury and Weltman~\cite{khoury}), an axionic coupling, a dilatonic coupling (see Damour and Donoghue~\cite{damour10}), etc\dots In this communication, we will focus on the microscopic modeling introduced by Damour and Donoghue~\cite{damour10} but the discussion that will follow can be translated into other types of coupling.

In the model introduced by Damour and Donoghue~\cite{damour10}, the matter part of the action is
\begin{equation}\label{eq:action}
	S_\mathrm{matter}\left[g_{\mu\nu},\Psi_i,\varphi\right]=\frac{1}{c}\int d^4x\sqrt{-g}\Big[\mathcal L_\mathrm{SM}\left(g_{\mu\nu},\Psi_i\right)+\mathcal L_\mathrm{int}(g_{\mu\nu},\varphi,\Psi_i)\Big]\, ,
\end{equation}
where $g$ is the determinant of the space-time metric $g_{\mu\nu}$, $c$ is the speed of light, $\mathcal L_\mathrm{SM}$ is the SM Lagrangian that depends on the SM fields $\Psi_i$ and $\mathcal L_\mathrm{int}$ parametrizes the interaction between matter and the scalar field. This interacting Lagrangian is parametrized through
\begin{equation}\label{eq:lag}
	\mathcal L_\mathrm{int}(g_{\mu\nu},\varphi,\Psi_i)\Big]=\varphi^i \Bigg[\frac{d_e^{(i)}}{4\mu_0}F_{\mu\nu}F^{\mu\nu}-\frac{d_g^{(i)}\beta_3}{2g_3}F_{\mu\nu}^AF^{\mu\nu}_A-\sum_{j=e,u,d}\left(d^{(i)}_{m_j}+\gamma_{m_j}d_g^{(i)}\right)m_j\bar \psi_j\psi_j\Bigg]\, ,
\end{equation}
with $F_{\mu\nu}$ being the standard electromagnetic Faraday tensor, $\mu_0$ the magnetic permeability, $F^A_{\mu\nu}$ the gluon strength tensor, $g_3$ the QCD gauge coupling, $\beta_3$ the $\beta$ function for the running of $g_3$, $m_j$ the mass of the fermions (electron and light quarks), $\gamma_{m_j}$ the anomalous dimension giving the energy running of the masses of the QCD coupled fermions and $\psi_j$ the fermion spinors.  The constants $d_j^{(i)}$ characterize the interaction between the scalar field $\varphi$ and the different matter sectors. Essentially two phenomenological modelings have been studied in the litterature: (i) a linear coupling~\cite{damour10,arvanitaki15} characterized by the coupling coefficients $d_j^{(1)}$ and (ii) a quadratic coupling~\cite{stadnik15,stadnik18} characterized by the coupling coefficients $d_j^{(2)}$.  Note that another convention (used e.g. by Stadnik and co-authors~\cite{stadnik15}) for the coupling coefficients is sometimes considered using dimensional $\Lambda^{(i)}_j$ coupling constants (see Appendix of Hees {\it et al}~\cite{hees18}).

This Lagrangian leads to the following effective dependency of five constants of Nature
\begin{eqnarray}
	\alpha_\mathrm{EM}(\varphi)&=&\alpha_\mathrm{EM}\left(1+d^{(i)}_e\varphi^i\right) \, , \label{eq:var1}\\
	m_j(\varphi)&=&m_j\left(1+d^{(i)}_{m_j}\varphi^i\right) \quad \textrm{for } j=e,u,d\,  \\
	\Lambda_3(\varphi)&=& \Lambda_3\left(1+d^{(i)}_{g}\varphi^i\right) \, ,\label{eq:var3}
\end{eqnarray}
where $\alpha_\mathrm{EM}$ is the electromagnetic fine structure constant, $m_j$ are the masses of the fermions  (the electron and the up and down quarks), $\Lambda_3$ is the QCD mass scale $\Lambda_3$ and the superscripts $^{(i)}$ indicate the type of coupling considered (linear for $i=1$ and quadratic for $i=2$). Note that, following Damour and Donoghue~\cite{damour10}, we introduce the mean quark mass $\hat m=\left(m_u+m_d\right)/2$ which depend also on the scalar field through~\footnote{The difference between the quark mass $\delta m$ is neglected in this communication, see Hees {\it el al}~\cite{hees18} for more details.}
\begin{equation}
	\hat m(\varphi)=\hat m\left(1+d^{(i)}_{\hat m}\varphi^i\right)\, \mathrm{with  }\, \,
 d^{(i)}_{\hat m}=\frac{m_u d^{(i)}_{m_u}+m_d d^{(i)}_{m_d}}{m_u+m_d}\, .
\end{equation}

In this communication, we will focus on free falling test masses and on frequency measurements. Both these systems need to be modeled from the microscopic Lagrangian presented above. 

\subsubsection{Free falling test masses}
Damour and Donoghue~\cite{damour10} have shown that the action describing matter including the microscopic interaction from Eq.~(\ref{eq:lag}) can be replaced at the macroscopic level by a standard point mass action, with each mass $A$ depending on the scalar field $m_A(\varphi)$. The effects produced by the scalar/matter coupling are totally encoded in the coupling function
\begin{equation}\label{eq:alpha}
	\alpha_A^{(i)}=\frac{\partial \ln m_A(\varphi)}{\partial \varphi^i}=d_g^{*(i)}+\bar\alpha_A^{(i)} \, ,
\end{equation}
where $d_g^{*(i)}$ is composition independent and $\bar\alpha_A^{(i)}$ is composition dependent. Both these coefficients depend on the matter/scalar coupling parameters $d_j^{(i)}$
\begin{eqnarray}
	d_g^{*(i)}&=& d_g^{(i)} + 0.093 \left(d_{\hat m}^{(i)}-d_g^{(i)}\right)+2.75\times 10^{-4}\left( d_{m_e}^{(i)}-d_g^{(i)}\right)+2.7\times 10^{-4}d_e^{(i)} \, , \\
	\bar\alpha_A^{(i)}&=&\left[Q_{\hat m}\right]_A\left(d_{\hat m}^{(i)}-d_g^{(i)}\right)+\left[Q'_{m_e}\right]_A \left(d_{ m_e}^{(i)}-d_g^{(i)}\right)+\left[Q'_{e}\right]_A d_e^{(i)}\, ,\label{eq:baralpha}
\end{eqnarray}
where the coefficients $\left[Q'_j\right]_A$ are the dilatonic charges for the body $A$ and characterize the sensitivity of the body $A$ to the scalar field. The values of these coefficients depend only on the composition of each body, they have been computed from theoretical atomic and nuclear calculations and their expression can be found in Damour and Donoghue~\cite{damour10}.
 
\subsubsection{Frequency measurements}
Frequency measurements (with atomic clocks or spectroscopy) are directly sensitive to a possible variation of the constants of Nature from Eqs.~(\ref{eq:var1}-\ref{eq:var3}). The effects produced by the scalar/matter coupling on a frequency measurement $\nu_C$ is encoded in the coupling function
\begin{equation}
	\kappa_C^{(i)}=\frac{\partial \ln \nu_C}{\partial \varphi^i}\, ,
\end{equation}
which is the equivalent of Eq.~(\ref{eq:alpha}) for test masses. This coefficient depends on the scalar/matter coupling parameters $d^{(i)}_j$ through
\begin{equation}
	\kappa_C^{(i)}=\left[k_e\right]_C d_e^{(i)} + \left[k_\mu\right]_C \left(d_{m_e}^{(i)}-d_g^{(i)}\right)+\left[k_q\right]_C \left(d_{\hat m}^{(i)}-d_g^{(i)}\right)\, ,\label{eq:kappa}
\end{equation}
where the coefficients $\left[k_j\right]_C$ are the sensitivity coefficients of the transition $\nu_C$ to the constants of Nature (e.g. $\left[k_e\right]_C=\partial \ln \nu_C/\partial \alpha_\mathrm{EM}$). These coefficients are similar to the dilatonic coefficients appearing in the modeling of test masses. The values of these coefficients depend only on the atomic properties of the frequency considered. They can be computed from theoretical calculations by solving numerically the Schr\"odinger equation (see e.g. the AMBiT software described in Kahl and Berengut~\cite{kahl}). Values of the $k_i$ sensitivity coefficients for different atomic transitions have been computed by Flambaum and collaborators~\cite{flambaum,dzuba}.

\subsection{Interpretation of three tests of the Einstein Equivalence Principle}\label{sec:EEP}
\subsubsection{Universality of Free Fall}
UFF tests consist in measuring the differential acceleration between two bodies $A$ and $B$ of different composition falling in the same gravitational potential. This differential acceleration is directly related to variation of the scalar field~\cite{hees18}
\begin{equation}
	\left[\Delta \bm a\right]_{A-B} = \bm a_A-\bm a_B=\left(\bar \alpha^{(i)}_B-\bar \alpha^{(i)}_A\right)\left[c^2 \bm \nabla\varphi^i +  \bm v\frac{d\varphi^i}{dt}\right] \, ,
\end{equation}
where the $\bar \alpha^{(i)}$ are given by Eq.~(\ref{eq:baralpha})~\footnote{Here, we assume the two bodies to be initially moving with the same velocity $\bm v_A=\bm v_B=\bm v$}. In order to design experiments that are the most sensitive to violation of the UFF, one needs to use two tests masses whose dilatonic coefficients are as different as possible and to locate the experiment in a region of space-time where variations of the scalar field are large.

\subsubsection{Local frequencies comparison}
One way to search for a violation of the local position invariance is to measure the frequency ratio between two frequencies $\nu_C$ and $\nu_D$ (atomic clocks or atomic lines observed with spectroscopic measurements) based on different atomic transitions and located at the same position. The observable is then $Y=\nu_C/\nu_D$ and its relative variation takes the form of
\begin{equation}
	\frac{Y}{Y_0}=K+\left(\kappa_{C}^{(i)}-\kappa_{D}^{(i)}\right)\varphi^i \, ,
\end{equation}
where the $\kappa^{(i)}$ are given by Eq. (\ref{eq:kappa}) and $K$ is a unobservable constant. In order to search for violations of the EEP, one needs to measure $Y$ at different space-time locations in order to be sensitive to scalar field variations. To design experiments that are sensitive to a violation of the EEP, one needs to use two frequency transitions whose sensitivity coefficients $k_j$ are as different as possible and monitor the ratio $Y$ in at least two different locations ideally characterized by large the scalar field differences.

\subsubsection{Redshift test}
In a typical redshift experiment, one measures the gravitational redshift between two clocks located in a different gravitational potential, this gravitational potential having been measured previously by using the motion of test masses (see e.g. Delva {\it el al}~\cite{delva}). First, let us consider a test mass $A$ moving in a gravitational potential. Its equation of motion is given by~\footnote{We assume the scalar field to be static and we keep only the leading order terms.}
\begin{equation}
	\bm a_A = \bm\nabla U -\alpha^{(i)}_A c^2 \bm\nabla \varphi^i = \bm\nabla \mathcal U_A \, , \, \textrm{where} \quad \mathcal U_A=U - \alpha^{(i)}_A c^2 \varphi^i \, ,
\end{equation}
where $U$ is the bare gravitational potential~\footnote{The bare gravitational potential is related to the low gravitational field expansion of the time component of the space-time metric appearing in the action from Eq.~(\ref{eq:action}): $g_{00}=-1+2U/c^2+\dots$.} and $\mathcal U_A$ is the observable gravitatonial potential as infered from the motion of the test mass $A$. Now, let us compare two clocks of the same type $C$ (this can be generalized easily) located in a different gravitational potential. The gravitational part of the redshift is given by
\begin{equation}
	\left[\frac{\Delta \nu}{\nu}\right]_\mathrm{grav}=\frac{U_\mathrm{em}}{c^2}+\kappa_C^{(i)}\varphi^i_\mathrm{em}-\frac{U_\mathrm{rec}}{c^2}+\kappa^{(i)}_C\varphi^i_\mathrm{rec} =\frac{\Delta\mathcal U_A}{c^2} + \left( \kappa^{(i)}_C+\alpha^{(i)}_A \right)\Delta \varphi^i \, ,
\end{equation}
where the subscript $em/rec$ refers to the emitter and receiver of the signal. In the last equality, we have replaced the bare gravitational potential by its observable counterpart (for a similar discussion, see Damour \cite{damour99}). The optimal measurement to search for a violation of the EEP with a redshift test is to compare two clocks with large sensitivity coefficients located in two regions characterized by a large scalar field difference.

\section{A model of ultralight bosonic Dark Matter}\label{sec:DM}
In this section, we will focus on a model where the scalar field is massive and plays the role of DM. This DM candidate is parametrized by the following action
\begin{equation}
	S=\int d^4x \frac{c^3\sqrt{-g}}{16\pi G}\left[R-2 g^{\mu\nu}\partial_\mu\varphi\partial_\nu\varphi-2\frac{c^2}{\hbar^2}m^2_\varphi\varphi^2 \right]+S_\mathrm{mat}\left[\Psi^i,g_{\mu\nu},\varphi\right]\, ,
\end{equation}
where $m_\varphi$ is the scalar field mass and the matter part of the action is given by Eq.~(\ref{eq:action}). At the cosmological level, the scalar field will oscillate at its Compton frequency~\cite{arvanitaki15,stadnik15} and on average will behave as a pressureless fluid, making it a perfect DM candidate. The amplitude of the scalar field oscillations $\varphi_0$ are directly related to the DM energy density through~\cite{arvanitaki15,stadnik15} $\varphi_0=(8\pi G \hbar^2 \rho_\mathrm{DM}/c^6m_P^2)^{1/2}\sim 7\times 10^{-31}$ eV$/m_\varphi$, where we used a local galactic value for the DM energy density of $\rho_\mathrm{DM}=0.4$ GeV/cm$^3$. 

For relatively low masses ($m_\varphi<10$ eV), the occupation number of the scalar field is high and $\varphi$ behaves as a classical field (see Derevianko~\cite{derevianko} for a detailed derivation). In addition, this model exhibits very nice galactic properties for masses at the level of $10^{-22}$ eV (see e.g. Marsh~\cite{marsh} or Hu {\it et al}~\cite{hu}). Finally, the DM velocity distribution in our Galaxy implies that the scalar field oscillations are coherent only over $10^6$ oscillations (see Derevianko~\cite{derevianko} for a detailed derivation), which impacts the data analysis of high frequency measurements for which the measurement time baseline is larger than the oscillations coherence time. One crude way to analyze data in that case consists in cutting the measurements in pieces shorter than the coherence time, but a better methodology can be developped. 

Interestingly, this model of DM will break the EEP and can be searched for in the lab. We will make the distinction here between two cases that produce very distinct phenomenologies: (i) a linear scalar/matter coupling and (ii) a quadratic scalar/matter coupling.

\subsection{A linear scalar/matter coupling}
\begin{figure}
\begin{minipage}{0.45\linewidth}
\centerline{\includegraphics[width=0.96\linewidth]{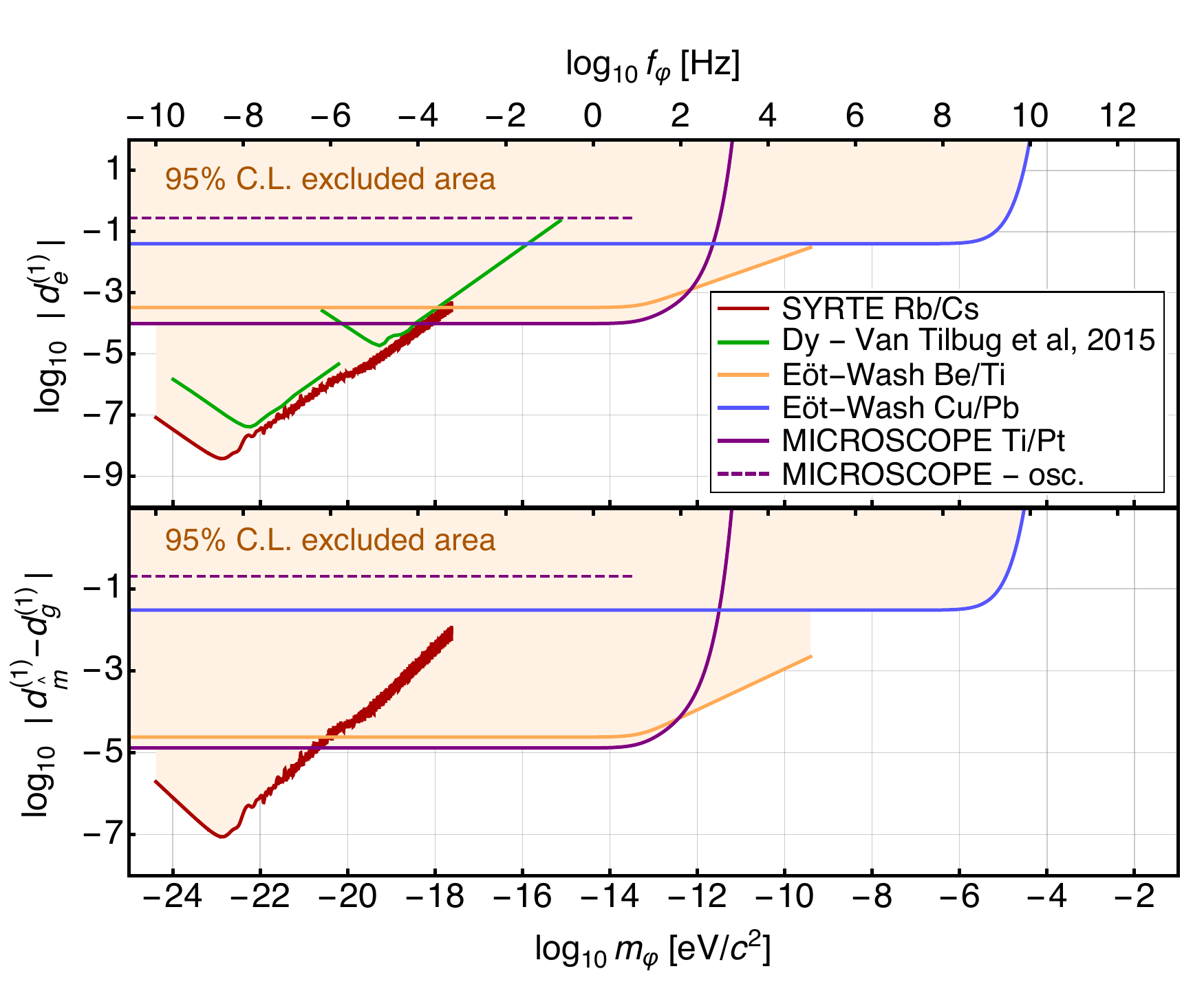}}
\end{minipage}
\begin{minipage}{0.5\linewidth}
\centerline{\includegraphics[width=0.96\linewidth]{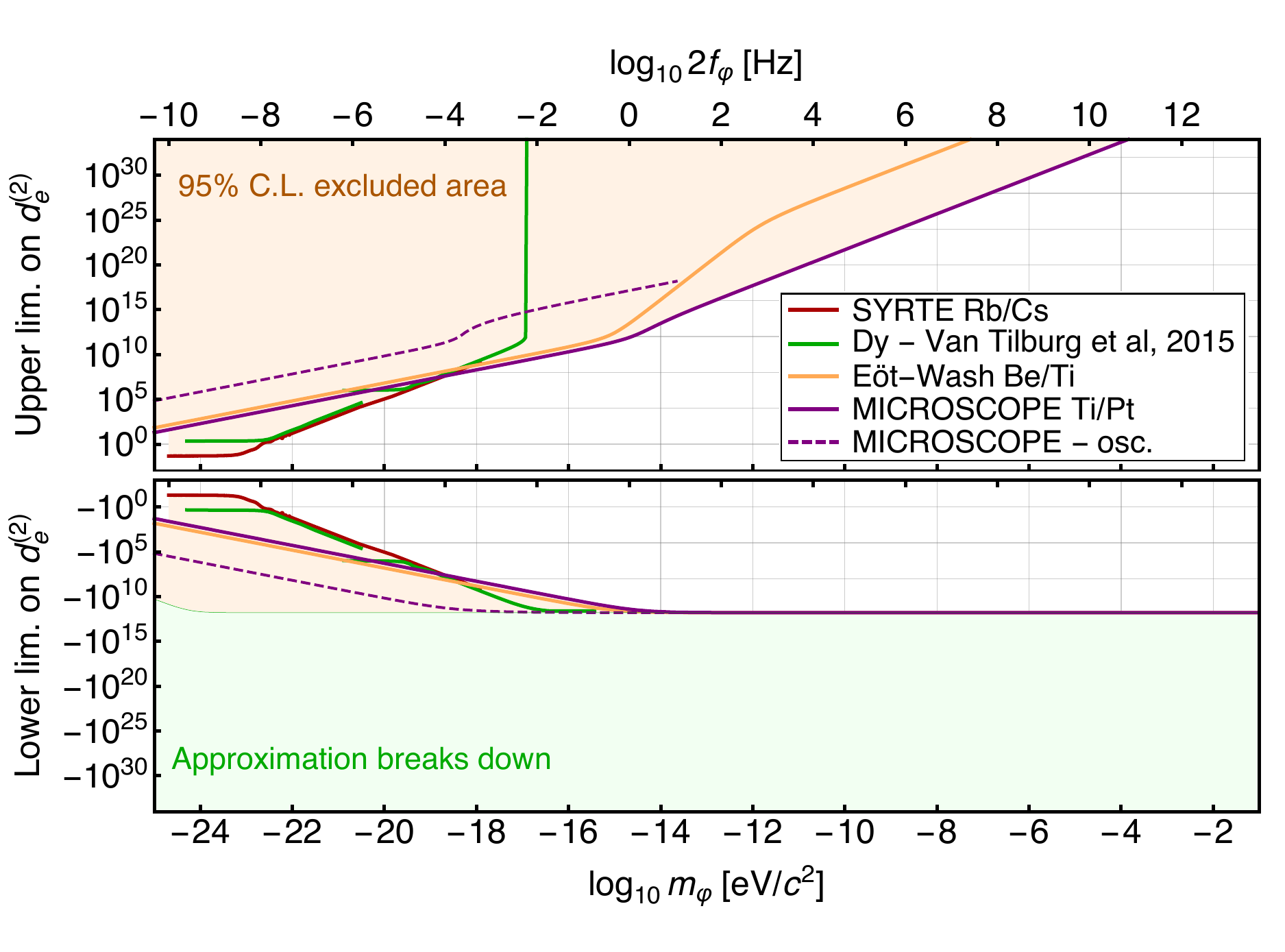}}
\end{minipage}
\caption[]{Constraints on the scalar/matter coupling  for a scalar DM model obtained from atomic clocks measurements~\cite{van-tilburg,hees16} and from UFF experiments~\cite{smith,schlamminger,microscope}. Left: constraints on the linear scalar/matter couplings $d^{(1)}_e$ and $d^{(1)}_{\hat m}-d^{(1)}_g$. Right: constraints on the quadratic scalar/matter coupling $d^{(2)}_e$. Similar figures for the other couplings can be found in Hees {\it et al}~\cite{hees18}.}
\label{fig:DM}
\end{figure}

In the linear case (i.e. setting $i=1$ in Eq.~(\ref{eq:lag})), the scalar field around a body $A$ is given by~\cite{hees18}
\begin{equation}
	\varphi^{(1)}=\varphi_0 \cos(\omega_\varphi t+\delta)- s_A^{(1)}\frac{GM_A}{c^2r}e^{-r/\lambda_\varphi} \, ,
\end{equation}
where $\omega_\varphi$ and $\lambda_\varphi$ are respectively the Compton pulsation and wavelength of the scalar field. The first term can be interpreted as DM. In that case the amplitude is directly related to the DM local energy density. This term can efficiently be searched for by using atomic clocks comparison~\cite{van-tilburg,hees16}. The second term is a Yukawa interaction and is generated by the central body. The scalar charge of the central body $s^{(1)}_A$ is proportional to the coefficient $\alpha_A^{(1)}$ from Eq.~(\ref{eq:alpha}) up to a geometric factor. UFF tests are particularly sensitive to this second term~\cite{berge}. Several measurements can be used to constrain such a model: (i) the comparison between Rb and Cs hyperfine transition frequencies from the dual atomic fountain from SYRTE~\cite{hees18}, (ii) comparison of two radio-frequency transitions using two isotopes of Dysprosium~\cite{van-tilburg}, (iii) UFF tests using torsion balances (Be versus Ti but also short distances test of Cu versus Pb)~\cite{smith,schlamminger} and (iv) the first result of the UFF test of the MICROSCOPE space-mission \cite{microscope}. The exclusion region for some of the linear scalar/matter coupling coefficients is presented on the left of Figure~\ref{fig:DM} and similar figures for the other couplings can be found in Hees {\it et al}~\cite{hees18}. For low scalar masses, atomic clocks provide the best constraints, large scalar masses are essentially constrained from lab UFF experiments while the MICROSCOPE result is competitive in the middle mass range.

\subsection{A quadratic scalar/matter coupling}
The case of a quadratic scalar/matter coupling (i.e. setting $i=2$ in Eq.~(\ref{eq:lag}) gives rise to a very rich phenomenology. Indeed, the scalar field solution is given by~\cite{hees18}
\begin{equation}
	\varphi=\varphi_0 \left[1-s^{(2)}_A\frac{GM_A}{c^2r}\right]\cos(\omega_\varphi t+\delta) \, ,
\end{equation}
where $s^{(2)}_A$ is the scalar charge of the central body, which depends non-linearly on $\alpha^{(2)}_A$ and on the body's compacity and $\varphi_0$ is related to the local DM energy density. First of all, it is worth to highlight the absence of any Yukawa interaction. Secondly, the amplitude of the scalar field oscillations are now depending on the location with respect to the central body, leading to a very rich and interesting phenomenology. In addition, the non-linearity characterizing $s^{(2)}_A$ leads to a screening mechanism for positive value of $\alpha_A^{(2)}$, meaning that the scalar field is screened close to the body and can become undetectable at its surface. On the other hand, negative values of  $\alpha_A^{(2)}$ lead to a scalarization mechanism where the scalar field is amplified~\cite{hees18}. The exclusion region for the quadratic scalar/matter coupling coefficient $d_e^{(2)}$ is presented on the right of Figure~\ref{fig:DM} and similar figures for the other quadratic couplings can be found in Hees {\it et al}~\cite{hees18}. For low scalar field masses, atomic clocks provide the best constraints while at larger masses, the MICROSCOPE result provides the most stringent constraints, due to the fact that being in space is favorable for such models. It is also intersting to note that the constraint is not the same for positive and negative values of $d^{(2)}_e$ due to the non-linearities.

\section{Can Sagitarius A$^*$ be a scalarized black hole?}\label{sec:EGB}
During last summer, the short-period star S0-2 (also named S2) experienced its closest approach from Sgr A*,  the SMBH in our GC. This event was followed closely by the UCLA Galactic Center Group \cite{do} and by the GRAVITY collaboration \cite{gravity}, which led to the detection of the relativistic contribution to the redshift of S0-2 at the level of $\sim 15 \%$~\cite{do,gravity}. If one parametrizes a deviation from the gravitational redshift~\footnote{Note that the gravitational redshift is only half the contribution of the relativistic redshift (the combination of the second order transverse Doppler and of the gravitational redshift) whose measurement is reported in~\cite{do,gravity}.} using the parametrization from Will~\cite{will} $\left[\Delta \nu/\nu\right]_\mathrm{grav}=(1+\alpha_\mathrm{red})U/c^2$, the GRAVITY result~\cite{gravity} writes $\alpha_\mathrm{red}=-0.2\pm 0.34$ while a similar result has been obtained by the UCLA group~\cite{do}. The discussion from Section~\ref{sec:EEP} shows that this result can be interpreted as a constraint on the coupling between a scalar field and matter (in this section, we will focus only on a static scalar field). To illustrate how these recent results can be used to constrain GR alternatives, let us consider the case of the quadratic Einstein-Gauss-Bonnet theory, recently studied by three different groups~\cite{antoniou,doneva,silva}. The action is given by
\begin{equation}
    S=\int d^4x \frac{c^3\sqrt{-g}}{16\pi G}\left[R-2g^{\mu\nu}\partial_\mu\varphi\partial_\nu\varphi+\frac{\eta}{8}\varphi^2\mathcal G\right]+S_\mathrm{mat}\left[g_{\mu\nu},\Psi_i,\varphi\right]\, ,
\end{equation}
where $\eta$ is a coupling parameter of dimension of length square that parametrizes the coupling between the scalar field and the Gauss-Bonnet invariant $\mathcal G=R^2-R^{\mu\nu}R_{\mu\nu}+R^{\mu\nu\alpha\beta}R_{\mu\nu\alpha\beta}$. One interesting feature from such theory relies in the existence of scalarized black holes solution of the vacuum field equations, in addition to standard GR black holes~\cite{antoniou,doneva,silva}. This means that there exists vacuum solution with non-trivial scalar field profile which can become large even at large distances from the horizon, a feature generated by the non trivial coupling between the scalar field and the Gauss-Bonnet curvature invariant. In particular, it can be shown that the scalar field profile at large distances from the horizon takes the form
\begin{equation}
    \varphi=\varphi_\infty + q\frac{GM}{c^2r} +\dots\, ,
\end{equation}  
where $\varphi_\infty$ is the asymptotic value of the scalar field and $q$ is the BH scalar charge which depends on the fundamental parameter $\eta$ and on the BH mass. This scalar charge can have values up to~\cite{doneva,silva} $q\sim 0.45$. If the scalar field is non-minimally coupled to matter, it will impact both the motion of S0-2 and the spectroscopic measurements, as described in Section~\ref{sec:EEP}. In particular, using the linear scalar/matter coupling ($i=1$ in Eq.~(\ref{eq:lag})), the Newtonian equation of motion for S0-2 writes
\begin{equation}
    \bm a_\mathrm{S0-2}=-\frac{GM}{r^3}\bm r(1-q\alpha^{(1)}_\mathrm{S0-2})=-\frac{G\mathcal M_\mathrm{S0-2}}{r^3}\bm r\, ,
\end{equation}
showing that the SMBH mass inferred from the motion of S0-2 ($\mathcal M_\mathrm{S0-2}$) is different from its bare mass $M$. In addition, the gravitational contribution to the redshift of S0-2 is given by~\footnote{The contribution from the frequency standard at reception is constant and is absorbed in the orbital fit.}
\begin{equation}
    \left[\frac{\Delta \nu}{\nu}\right]_\mathrm{grav, S0-2}=\frac{GM}{c^2r}\left(1+\kappa^{(1)}_{S0-2}q\right)=\frac{G\mathcal M_\mathrm{S0-2}}{c^2r}\left[1+q \left(\kappa^{(1)}_\mathrm{S0-2}+\alpha^{(1)}_\mathrm{S0-2}\right)\right] \, .
\end{equation}
This shows that for this class of theory, the gravitational redshift parameter can directly be mapped to the fundamental parameters of the theory (i.e. the constant $\eta$ and the scalar/matter coupling coefficients) through $\alpha_\mathrm{red}=q\left(\kappa^{(1)}_\mathrm{S0-2}+\alpha^{(1)}_\mathrm{S0-2}\right)$ where $\alpha^{(1)}_\mathrm{S0-2}$ depends on the composition of S0-2 and is given by Eq.~(\ref{eq:alpha}) and $\kappa^{(1)}_\mathrm{S0-2}$ depends on the atomic property of the Br-$\gamma$ atomic line used to measured S0-2's radial velocity and is given by Eq.~(\ref{eq:kappa}). The full derivation of the exclusion region within the parameters space $\left(\eta, d^{(1)}_i\right)$ is currently a work in progress. 

\section{Conclusion}\label{sec:conclusion}
Tensor-scalar theories of gravitation remain a wide class of alternatives to GR with various motivations ranging from higher dimension scenarios to DM or dark energy model. In general, one expects the scalar field to couple non-universally with matter (unless this is prevented by some kind of symmetry) which leads to a violation of the EEP. In this communication, we reviewed some impacts induced by a violation of the EEP by a scalar field and how such signatures can be constrained by various types of experiments ranging from laboratory experiment (atomic clocks comparisons and torsion balances), to space-mission (with MICROSCOPE) to astrophysical measurements (like e.g. in our GC). It is likely that the search for a breaking of the EEP induced by scalar fields will be pursued and improved in the future with improved experiments, with new types of experiments and in regimes under-explored (or not exploread at all) so far.

\section*{Acknowledgments}
AH and ES thank the organizers for financial support to attend the conference and AH thanks A. F\"uzfa for interesting discussions on the subject.

\end{document}